\documentclass[12pt]{article}
\usepackage[english]{babel}
\usepackage{pdfpages}
\usepackage{lmodern}
\usepackage{graphicx} 
\usepackage{wrapfig} 
\usepackage{textcomp} 
\usepackage[utf8]{inputenc}
\usepackage[T1]{fontenc}     
\usepackage{microtype}       
\usepackage[scaled]{helvet} 
\usepackage{setspace} 
\usepackage{dsfont}
\usepackage{authblk}

\DeclareSymbolFont{letters}{OML}{ztmcm}{m}{it}
\usepackage{color}
\usepackage{bm}
\usepackage{bigdelim, multirow}

\usepackage{array}
\usepackage{listings}
\usepackage{adjustbox}
\usepackage{subcaption}
\usepackage{multirow} 
\usepackage[OMLmathrm,OMLmathbf]{isomath}
\usepackage{booktabs} 
\usepackage[hang,flushmargin]{footmisc}
\usepackage{pdflscape}
\usepackage{comment}
\usepackage{capt-of}

\usepackage[backend=biber, firstinits=false,uniquename=false,style=authoryear, citestyle=authoryear, maxnames=10, maxbibnames=10, natbib=true]{biblatex} 

\usepackage[T1]{fontenc} 
\usepackage[margin=1.5in]{geometry}
\usepackage{float}
\usepackage{amsmath}
\usepackage{algorithm}
\usepackage{algpseudocode}
\usepackage{etoolbox} 
\usepackage{amsthm}
\usepackage{amsfonts}
\usepackage{amssymb}
\usepackage[section]{placeins}
\usepackage{diagbox} 
\usepackage{hyperref}
\usepackage{graphicx} 
\usepackage{fixmath} 
\usepackage{tikz}
\usepackage{footnote}
\makesavenoteenv{tabular}
\makesavenoteenv{table}
\usepackage{setspace}
\usepackage{multicol}

\usepackage{caption}
\usepackage{graphicx} 
\usepackage{amsmath}

\begin{document}


\title{Using LASSO for Variable Selection in Exponential Random Graph models}

\author[1]{Sergio Buttazzo}
\author[1]{Göran Kauermann}
\affil[1]{LMU Munich}
\date{}
\maketitle

\begin{abstract}
\begin{spacing}{1}
    The paper demonstrates the use of LASSO-based estimation in network models. Taking the Exponential Random Graph Model (ERGM) as a flexible and widely used model for network data analysis, the paper focuses on the question of how to specify the (sufficient) statistics, that define the model structure. This includes both, endogenous network statistics (e.g. degree distribution, edgewise shared partners etc.) as well as statistics involving exogenous covariates on the node level. LASSO estimation is a penalized estimation that shrinks some of the parameter estimates to be equal to zero. As such it allows for model selection by modifying the amount of penalty. The concept is well established in standard regression and we demonstrate its usage in network data analysis, with the advantage of 'automatically' providing a model selection framework.  
\end{spacing}
\end{abstract}


\section{Introduction}
The analysis of network data with exponential random graph models (ERGM) has become very popular in recent years. Originally proposed by \citet{holland1981} and \citet{FraStr1986}, the model classed gained interest with stable fitting procedures proposed and discussed in 
\citet{Sni2002}, \citet{HunKriSch2012} and \citet{HumHunHan2012}, and implemented ready to use software, see \citet{hunter2008ergm} and \citet{HanHunBut2021}. We refer to \citet{Kol2009} and \citet{KolCsa2014} for a general introduction to network data analysis, see also \citet{LusKosRob2012} or \citet{harris2013introduction} for a focus on ERGMs or more generally \citet{borgatti2018analyzing} and \citet{cranmer2020inferential}. The model class has been extended in various ways by including random components (\citealp{kevork2022bipartite} or \citealp{thiemichen2016bayesian}), temporal aspects (\citealp{leifeld2019theoretical} or \citealp{leifeld2018temporal}) or mixtures of models (\citealp{schweinberger2018hergm}), to mention a few ideas. Generally, the core of an ERGM is the specification of sufficient network statistics that define the model structure and allow for interpretations regarding the network formation.  


A network consists of a set of nodes $\mathcal{N}$ and a set of edges $E\subset \mathcal{N} \times \mathcal{N}$, which can be expressed as a network related adjacency matrix  $Y \in \{0,1\}^{N\times N}$ where $Y_{ij} = 1 \Leftrightarrow (i,j) \in E$, i.e. if there is an edge from node $i$ to node $j$.  
Assuming undirected edges, meaning that network ties do not contain information about the orientation of the tie, we have $Y_{ij} = Y_{ji}$, i.e.\ the adjacency matrix $Y$ is symmetric.  Self-loops are typically left undefined, meaning that the diagonal of $Y$ is not specified. With $N = |\mathcal{N}|$ we define the number of nodes in the network. In this paper, we will restrict ourselves to undirected networks, even though all proposed methods in this paper can be extended readily to directed ties.

The general structure of an ERGM is of exponential family type, i.e.\
the distribution of the random matrix $Y$ is given by:
\begin{align}
\label{eq:ergm}
    P(Y=y ; \theta) = \exp\{ \textbf{s}(y)^\top\theta - \kappa(\theta)\}.
\end{align}
Here $\theta \in \Theta$ is the parameter of interest and $\kappa(\theta)$ is a log-normalization constant. 
The vector $\textbf{s}(y)= (s_0(y), s_1(y), s_2(y), \cdots , s_p(y))^\top$ contains the model statistics expressing endogenous network effects. The first component $s_0(y) =  \sum_{i=1}^{n} \sum_{j>i} y_{ij}$ takes the role of the ``intercept'' in ERGMs. The remaining components take the role of model variables. A number of different reasonable structural statistics have been proposed in \citet{SniPatRobHan2006} and \citet{ergmterms}, see also \citet{snijders2011statistical}. The model can easily be extended to also include exogenous covariates $x$, which results by constructing network statistics $\textbf{s}(y,x)$ that depend on both, the network as well as tie or node-specific quantities $x$. For ease of notation, we omit this generalization for the moment.

In general, the concrete choice of the statistics typically depends on the research questions one wants to focus on. In other words, quite commonly in network data analysis, one selects the set of statistics based on substance matter research questions and then estimates the specified model and interprets the resulting parameter estimates. This approach has a couple of pitfalls. First, many of the commonly used statistics are heavily correlated, as they measure similar properties of the network. This leads to excessively high running times and poor model performances when used in combination with each other. Secondly, the model statistics need to be specified in advance, which requires some level of expertise by the researcher. Third, overall, model evaluation of the fitted model is necessary but it might be expedient to have model selection and model estimation somewhat combined. 
%


 In this paper, we propose a data-driven approach to select the statistics in $\textbf{s}(y)$. 
 To do so we make use of LASSO, where the acronym stands for {\bf L}east {\bf A}bsolute {\bf S}hrinkage to obtain a model {\bf S}election {\bf O}perator, which allows for data-driven model selection. To do so we start with a rich selection, that is we include numerous statistics in the statistics vector $\textbf{s}(y)$ so that its dimension $p$ is large. This in turn requires inducing a prior structure on the parameter in terms of penalization to make the model estimable. Originally proposed in \citet{tibshirani1996regression}, see also \citet{santosa1986linear}, the method has gained wide and still increasing interest in statistics, data analysis, and machine learning, see exemplary \citet{zhao2006model}. A Bayesian formulation of the LASSO approach is provided in \citet{park2008bayesian}. General properties of the LASSO when applied to OLS regression have been extensively studied. \citet{knightfu} show consistency and asymptotic normality for the LASSO estimator, provided appropriate convergence of the amount of penalization. \citet{zhao2006model} provide a necessary and sufficient condition for the LASSO estimator to achieve consistent selection (i.e. selecting the true parameters).

 Interestingly enough, variable selection and the development of model quality measures in general, even though having huge importance when dealing with ERGMs, have received little academic attention compared to other issues connected with the statistical modeling of network data. An example is given by \citet{stepwiseERGM}, who introduce and discuss a bottom-up, stepwise variable selection algorithm based on relative AIC change. We also refer to \citet{hunter2008goodness} who propose to assess an ERGM's performance by directly comparing the observed network with a sample from the model, using a set of ad-hoc chosen statistics as a judgment meter. LASSO itself has been applied to ERGMs by \citet{bayesianlasso}, even though their work is focused on an entirely Bayesian point of view. Generally, though, while LASSO estimation is widely used in a classical regression setup, its application in ERGMs is certainly less developed. Some first ideas have been proposed for instance in \citet{betancourt2017bayesian} or in recent dissertations, see \citet{ren2023inferring}, \citet{modisette2023penalized}, \citet{betancourt2015doctor}. In this paper, we elaborate this further and focus on the use of LASSO estimation for model selection of ERGMs. It is noteworthy that the parameters given by the LASSO approach are obtained through the introduction of a bias, that leads to decreasing performances in terms of model fit. Instead, we pursue to make use of the LASSO to obtain a ranking of importance of the model variables and use this ranking as a guideline regarding which variables to include in a new model, fit classically without penalization. More details are provided in the paper.


The paper is structured as follows: in Section 2 we define the mathematical foundations upon which the LASSO variable selector is built. First we explicitly define the ERGMs by characterizing the probability measure that defines this class of models. Additionally, we present the set of network statistics that are commonly used as model variables and discuss parameter estimation. Subsequently, we switch the focus on a penalized version of the ERGM likelihood function and derive a variable ranking algorithm based on penalized parameter estimation. In Section 3 we discuss applications of the above-discussed method, by presenting results regarding a number of simulated networks and two real-life (undirected, binary) networks. Finally, in Section 4, we provide a discussion on potential future research questions linked with the topic at hand.

\section{LASSO Estimation in ERGMs}

\subsection{Parameter estimation in ERGMs}\label{ergm_est}
We start our presentation with a short review of parameter estimation in the ERGM (\ref{eq:ergm}). This will ease the presentation of LASSO estimation subsequently.

The problematic component in (\ref{eq:ergm}) is given by the normalizing constant
\begin{align}
    \label{eq:kappa}
\exp\left(\kappa\left(\theta\right)\right) = \sum_{y\in\mathcal{Y}} \textbf{s}(y)^\top\theta
\end{align}
where the sum is over all possible network formations with $N$ nodes. Calculation of (\ref{eq:kappa}) is numerically infeasible unless for miniature networks, but the obstacle can be overcome by Markov Chain Monte Carlo (MCMC) methods (\citealp{mcmc1}). The idea of the approach is to estimate the parameters iteratively, relying on samples from a temporary version of the model at each step. Markov chain-based methods are particularly useful in the case of ERGMs, since the likelihood function is only needed up to a normalization constant: a sample can be obtained without having to directly evaluate the constant $\kappa(\theta)$. See \citet{Sni2002}, \citet{hunter2006}  or \citet{HunKriSch2012} for further details.

If the normalization constant were tractable, the MLE could be directly computed by the Newton-Raphson algorithm with the iteration step:
\[
\theta^{(t+1)} = \theta^{(t)} - \Sigma_{\theta^{(t)}}\left(\textbf{s}(Y)\right)^{-1}\left(\mathbb{E}_{\theta^{(t)}}[\textbf{s}(Y)] - \textbf{s}(y)\right)
\]
where $\Sigma_{\theta^{(t)}}\left(\textbf{s}(Y)\right)$ is the covariance matrix of the statistics vector $\textbf{s}(Y)$ at iteration step $n$. The quantities $\Sigma_{\theta^{(t)}}\left(\textbf{s}(Y)\right)$ and $\mathbb{E}_{\theta^{(t)}}[\textbf{s}(Y)]$ are not directly computable but, assuming an effective sampling algorithm, many variations are possible for parameter estimation. 
The approach of \citet{crouch}, based on \citet{mcmc1}, is to use a sample from the distribution with parameter $\theta^{(t)}$ to get an approximation of the moment equation in a neighbourhood of $\theta^{(t)}$ and solve it to get the new estimate $\theta^{(t+1)}$. An alternative approach, given by \citet{Sni2002}, is to make use of the \citet{robbinsmonro} stochastic iterative algorithm, and obtain an approximate value for $\Sigma_{\theta^{(n)}}\left(\textbf{s}(Y)\right)$ and $\mathbb{E}_{\theta^{(t)}}[\textbf{s}(Y)]$ through simulation:

\begin{align*}
    \mathbb{E}_{\theta^{(t)}}[\textbf{s}(Y)] &\approx \frac{1}{M}\sum_{i=1}^M \textbf{s}(y_i) = \bar{\textbf{s}}\\
    \Sigma_{\theta^{(t)}}\left(\textbf{s}(Y)\right) &\approx \frac{1}{M}\sum_{i=1}^M \left(\textbf{s}(y_i) - \bar{\textbf{s}}\right)\left(\textbf{s}(y_i) - \bar{\textbf{s}}\right)^\top
\end{align*}
The approximation for the covariance, however, can lead to singularities in the iteration process when $M$ is small. On the other hand, due to the sampling process being time-consuming, it would be preferable to not have to rely on a large number of simulations in every step. \citet{Sni2002} tackles the issue by estimating $D = \Sigma_{\theta^{(t)}}\left(\textbf{s}(Y)\right)$ only for $t=1$ and keeping it constant throughout the entire iteration. Modern works on stochastic optimization (see \citealp{bottoutricks}, \citealp{adam}, \citealp{adadelta}) show that the evaluation of second-order derivatives of the objective function is not needed, and convergence can be achieved by choosing an appropriate scalar learning rate sequence. We direct the reader to \citet{bottoutricks} for additional details and limit ourselves to using the suggested practical sequence of $\eta_t \sim 1 / \sqrt{t}$.
The resulting algorithm for ML estimation in ERGMs is sketched in Algorithm \ref{alg:sgd} below.

\begin{algorithm}
\caption{Simulation-based SGD maximum likelihood estimation}\label{alg:sgd}
\begin{itemize}
\item Assume an observed network $y$ and an ERGM with chosen model statistics $\mathbf{s}(\hat{y})$. Set $\theta = \theta^{(0)}$, $M\ge 0$ and a sequence of learning rates $\eta_t > 0$, such that $\eta_t = O(t^{-\frac{1}{2}})$.\\
\item For $t \ge 0$, cycle until convergence:
    \begin{itemize}
    \item Using algorithm 1, draw an $M$-size sample $\{y_i^t,\dots,y_M^t\}$ from the model, with parameters $\theta = \theta^{(t)}$;
    \item Evaluate $\Delta_t = \frac{1}{M}\sum_{i=1}^M \mathbf{s}(\hat{y}) - \mathbf{s}(y_i^t)$;
    \item Update $\theta^{(t+1)} = \theta^{(t)} + \eta_t\cdot\Delta_t$
    \end{itemize}
\end{itemize}
\end{algorithm}

\bigskip

\subsection{Penalization}

Let us now extend the above to LASSO estimation. 
To incorporate  LASSO approach to \textit{penalized} maximum likelihood estimator in ERGMs we add a penalty term to the log likelihood function and obtain the penalized log likelihood

\begin{equation}\label{penalizedergm}
        \ell_{pen}(Y=y ; \theta, \lambda) =  \textbf{s(y)}^\top\theta - \kappa(\theta) - \lambda\cdot\|\theta'\|_1.
\end{equation}
It needs to be noted that the edges statistic takes the role of the intercept in the case of ERGMs and as such its parameter should remain unpenalized. With this goal in mind, in (\ref{penalizedergm}), $\theta'_j = \theta_j$ for each $j\neq 1$, while $\theta'_1 = 0$.
Due to the penalizing term having a singularity at $\theta = 0$, simply applying the Stochastic Gradient Descent (SGD) iteration from Algorithm \ref{alg:sgd} to the log-likelihood in (\ref{penalizedergm}) would result in excessive oscillations when approaching any $\theta$ having one or more components $\theta_j = 0$. Since the goal of the LASSO approach is exactly to select a parameter vector containing null components, the update rule needs to be adapted.

The research output on gradient-based optimization methods applied to penalized functions is very vast. An overview is given in \citet{bach} and a selection of elaborate approaches can be found, for example, in \citet{frowardbackwardsplitting} and \citet{truncgrad}. The classic approach, which we follow here, is the method of subgradients. A subgradient of a function $f(\textbf{x})$ at point $x_0$ is defined as a vector $\textbf{v}$ such that

\[
f(\mathbf{x}) - f(\mathbf{x}_0) \ge \mathbf{v}\left(\mathbf{x} - \mathbf{x}_0\right).
\]
This definition is a generalization of the concept of gradient and is applicable in cases where the latter is not defined, such as the LASSO penalizing term. In the specific case we are interested in, a suitable choice of subgradient for the function $f(\theta) = \lambda\cdot\|\theta\|_{1}$ is 

\[
\mathbf{v} = \lambda\cdot sign(\theta)
\]
where $sign(\theta)$ is equal to $+1$ for positive components of $\theta$, $-1$ for negative components and $0$ for null components.

\bigskip

The penalized maximum likelihood estimation algorithm we propose can be summarized as shown in Algorithm \ref{alg:sgd2}. Given an observed network, we choose a large set of candidate model statistics, a penalizing factor, a starting parameter vector and an appropriate learning rate sequence. At each step, a sample of networks is drawn from the penalized likelihood with the current parameter, which is used to obtain an estimation of the subgradient of the log-likelihood. Finally, the parameter vector is updated by SGD, using the appropriate learning rate and the subgradient estimation previously obtained.

\bigskip

\begin{algorithm}
\caption{LASSO variable selector}\label{alg:sgd2}
\begin{itemize}
\item Assume an observed network $y$ and an ERGM with chosen model statistics $\mathbf{s}(y)$. Set $\theta = \theta^{(0)}$, $M\ge 0$, penalization term $\lambda>0$ and a sequence of learning rates $\eta_t > 0$, such that $\eta_t = O(t^{-\frac{1}{2}})$.
\item For $t \ge 0$, cycle until convergence:
    \begin{itemize}
    \item Draw an $M$-size sample $\{y_i^t,\dots,y_M^t\}$ from the model, with parameters $\theta = \theta^{(t)}$;
    \item Evaluate $\Delta_t = \frac{1}{M}\sum_{i=1}^M \mathbf{s}(y) - \mathbf{s}(y_i^t)$;
    \item Update $\theta^{(t+1)} = \theta^{(t)} + \eta_t\cdot\left(\Delta_t - \lambda\cdot sign\left(\theta^{(t)}\right)\right)$
    \end{itemize}
\item Obtain converged vector of parameters $\hat{\theta}(\lambda)$
\item Select the vector of statistics $\{s_i\}$ with  $\hat{\theta}_{i}(\lambda) \neq 0$.
\end{itemize}
\end{algorithm}

\bigskip

The vector of statistics $s_i(y)$ obtained through Algorithm \ref{alg:sgd2}  depends strongly on the choice of the penalizing term $\lambda$. The higher the degree of penalization, the lower the number of variables selected. 

\subsection{Variable Importance}

Since the set of parameter estimates obtained by Algorithm \ref{alg:sgd2} is penalized (and, as a consequence, biased), it is not recommended to use the output as final model estimates. However,  we argue that relevant information is contained in the penalizing term itself and it can be used to determine the importance of a model statistic.  Along this line of arguments  we define the \textit{importance score} for the statistic $s_i(y)$ as the maximum value of $\lambda$ for which the penalized estimated parameter $\hat{\theta}_{i}(\lambda)$ is non-zero:
\[
R_i = \max \{\lambda\ |\ \hat{\theta}_{i}(\lambda) \neq 0\}.
\]
The larger the value $R_i$, the more penalization is needed to set the parameter estimate to zero and hence the more important is the corresponding statistic $s_i(y)$. 
The suggested use for the LASSO selector is therefore the following: 
\begin{itemize}
    \item Run the algorithm for different values of $\lambda$, in order to obtain a ranking of variables based on their importance score;
    \item Choose a threshold value $\lambda_{threshold} $  for the importance score, so that all variables with $R_i > \lambda_{threshold}$ are included in the model.
    \item Fit an unpenalized ERGM, using the variables selected in the previous step.
\end{itemize}

It is worth noting that above, no indication is given on how to choose the threshold value $\lambda_{threshold}$. One possibility is to make use of the Akaike Information Criterion (AIC), meaning that the threshold is obtained such that adding the next variable in the ranked order of the statistics would result in a lower AIC value. Another approach is to follow p-values and choose the threshold such that adding a single new statistic based on the ranking obtained through $R_i$ would result in non-significant estimates. Whatever method one uses to determine the threshold $\lambda_{threshold}$, the importance ranking based on the LASSO estimate reduces the set of candidate models for model selection considerably.
We can conclude that the main appeal of the LASSO selector is to provide a ranking of statistics importance.

\subsection{Standardization of network features}

Even in classical i.i.d. models, the use of penalized estimation often requires to standardize the component variables, due to their different range of possible values: the size of the effect of a given variable, namely its parameter, depends on the magnitude of values taken by the variable itself. When penalization is applied to a set of parameters, it is necessary for their numerical values to be comparable. A commonly used way to deal with this problem is to standardize the variables, i.e. rescale the range of values taken by each variable to a fixed interval.

When it comes to network data the issue becomes slightly more tricky. Since we are provided with a single network observation, for variables such as structural statistics we are not provided with the sample standard deviation in the usual way. Instead, a larger sample is to be generated from a model of the observed network. In our case, we opted for a simple Erd\H os-Rényi model, as the goal here is not precision in the fit, but rather an estimation of the magnitude of structural variables \textit{per se}. The estimation is provided by the observed values in the sample: given a component $s_i(y)$ of the network statistics vector, and a sample $\{y_1, \dots, y_M\}$ of networks simulated from an Erd\H os-Rényi model, the scaled statistic will be:

\begin{equation}\label{standardization}
    s_{i,scaled}(y) = SD\left(s_i(y_1),\dots, s_i(y_M)\right)^{-1} s_i(y)
\end{equation}
where the standard deviation is defined as usual for an $M$-sized sample as 

\[
SD\left(s_i(y_1),\dots, s_i(y_M)\right) = \sqrt{\frac{1}{M-1}\sum_{j=1}^M \left(s_i(y_j) - \bar{s}_i\right)^2}.
\]

The same methodology also applies to nodal covariates: in ERGMs nodal covariates are treated as a single summary statistic relative to the network and are, therefore, not different in nature from structural statistics. 

In Section \ref{ergm_est} we discussed the problems with variance estimation when it comes to simulations: in particular we discussed that a large sample size is needed to avoid singularities, and how a large sample size can put a strain on running times when samples need to be drawn in each step of an iterative algorithm. In this case this is not a limiting issue: the variance estimation used for standardization is performed one-off at the start of the cycle and not repeated in each step of the iteration.

\section{Simulation Studies}
\subsection{Endogenous Network Statistics}

Before presenting the estimation and model selection strategy on real-world networks we want to explore the procedure in simulations. These simulations allow us to study whether LASSO can be used for model selection in ERGMs. To do so, we run simulations with known model setups which are described subsequently. Traditionally the \texttt{triangle} count has been used as a meaningful statistic to measure triadic closure for networks, while the \texttt{2-star} count as a meaningful statistic in order to measure triadic non-closure. However, literature has shown (\citealp{degeneracy}) that these two particular statistics increase the chances for the final model to be \textit{near-degenerate}. This means that the distribution resulting from the model will likely have a double peak shape and assign unreasonably high probabilities to near-empty or near-complete networks. A more favourable choice of variables uses a geometrically weighted sum of edgewise shared partners (\texttt{gwesp}) as a substitute for the triangle count, and a geometrically weighted sum of non-edgewise shared partners (\texttt{gwnsp}) as a substitute for the 2-star count.

These two statistics are formally defined as follows:

\begin{align*}
s_{gwesp}(y,\alpha) = e^\alpha \sum_1^{n-2}\left(1-\left(1-e^{-\alpha}\right)^i\right)esp(i)\\
s_{gwnsp}(y,\alpha) = e^\alpha \sum_1^{n-2}\left(1-\left(1-e^{-\alpha}\right)^i\right)nsp(i).
\end{align*}
Here $esp(i)$ and $nsp(i)$ count the number of edges (or non-edges, respectively) that share exactly $i$ nodes as neighbours and $\alpha$ is a decay parameter which can be set manually, as in our case, or estimated through the use of a \textit{curved} exponential family distribution (\citealp{curvedergm}).

Our first simulation study involves 3 different setups: in each setup a sample of networks is drawn from a known ERGM distribution and a run through the LASSO variable selector, in order to test if the procedure is able to select the variable used in the original distribution. In every setup we utilize 20 simulation replicas, namely 20 different simulated networks. The first three setups are performed on networks featuring $N=50$ nodes. In setup number 1 networks are drawn from an ERGM distribution using \texttt{edges} and \texttt{gwesp} as variables, with coefficients $\theta_{edges} = -3.5$ and $\theta_{gwesp} = 1.0$. In setup number 2 the sample is drawn from an ERGM distribution using \texttt{edges} and \texttt{gwnsp} as variables, with coefficients $\theta_{edges} = -2.0$ and $\theta_{gwesp} = 0.5$. In setup number 3 the sample is drawn from an Erd\H os-R\'enyi model, equivalent to an ERGM using only \texttt{edges} as a variable, with coefficient $\theta_{edges} = -1.5$. The 3 simulation setups are then repeated with networks featuring $N=150$ nodes. In all setups the starting set for the selecting procedure contains \texttt{gwesp}, \texttt{gwnsp}, and \texttt{gwdegree} as potential candidates. 

The study is conducted as follows: in each setup and for each sampled network, the $\theta$ coefficients are evaluated for different magnitudes of the penalization parameter $\lambda$. Afterwards, importance score $R_i$, i.e. the highest $\lambda$ for which the coefficient is non-zero is recorded for each variable. It must be noted that the importance score obtained depends on the sequence of penalization that is used.
We use the R package \textit{ERGM} from the Statnet project for all simulations (\citealp{HanHunBut2021}).

\begin{figure}
\centering
\includegraphics[
  height=7cm,
  width=15cm]{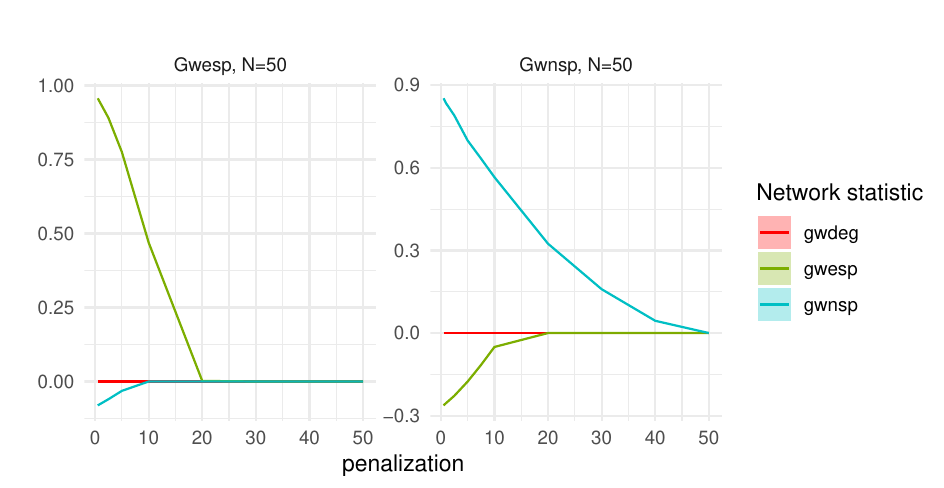}
\caption{Example of coefficients path in setup 1 (left), and setup 2 (right).}
\label{example1_plots}
\end{figure}

In Figure \ref{example1_plots} we show an example run on two networks from setup 1 and 2, respectively, and show how the coefficients of each variable change according to the magnitude of the penalization applied: setup 1 uses networks generated from an ERGM using \texttt{gwesp} as model statistic and, as such, the \texttt{gwesp} term is selected in the first example, with an importance score of 20. On the other hand, setup 2 uses networks generated from an ERGM using \texttt{gwnsp} as model statistic. This is again accurately reflected in the example, where \texttt{gwnsp} is selected with an importance score of 50.

\begin{figure}
\centering
\includegraphics[width=1.2\textwidth,
height=1.2\textheight,keepaspectratio]{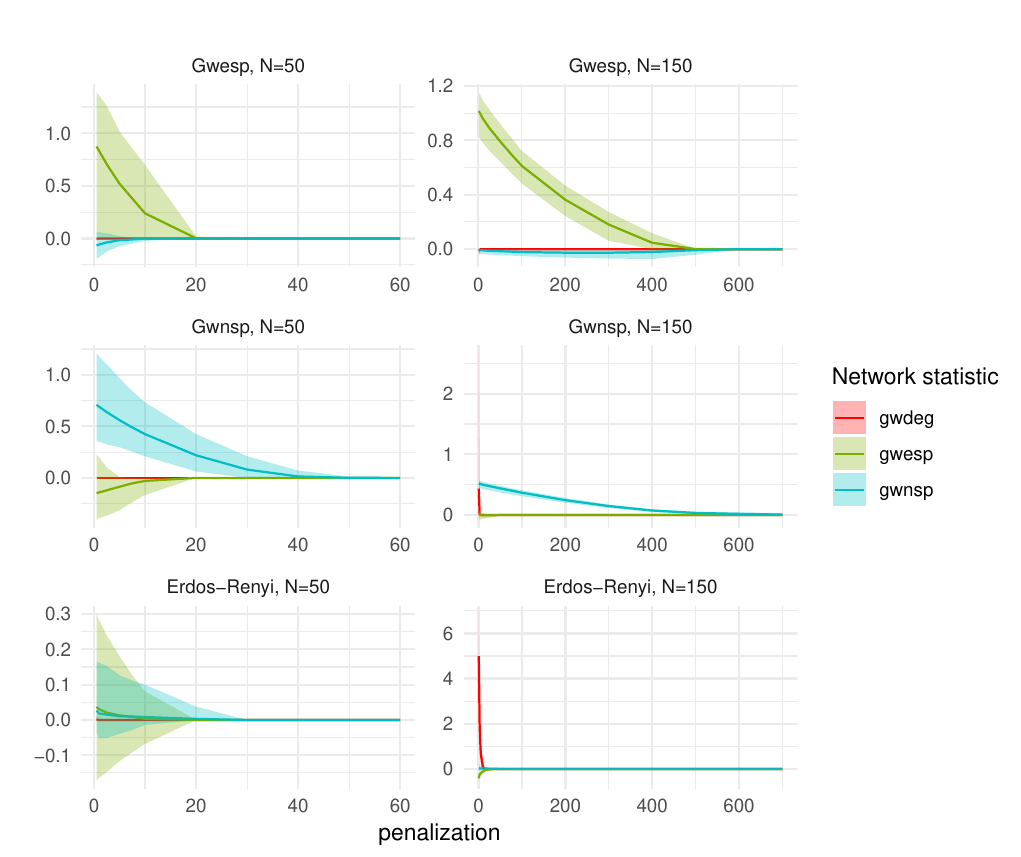}
\caption{Coefficient path in the different simulation setups, for different values of the penalization parameter. The first row corresponds to setup 1, the second to setup 2, the third to setup 3. Left column corresponds to network size $N=50$, right column to network size $N=150$. The colored line represents the average across the simulated networks while the coloured area covers the maximum and minimum.}
\label{simul_plot}
\end{figure}

\begin{figure}
\centering
\includegraphics[width=\textwidth,
height=\textheight,keepaspectratio]{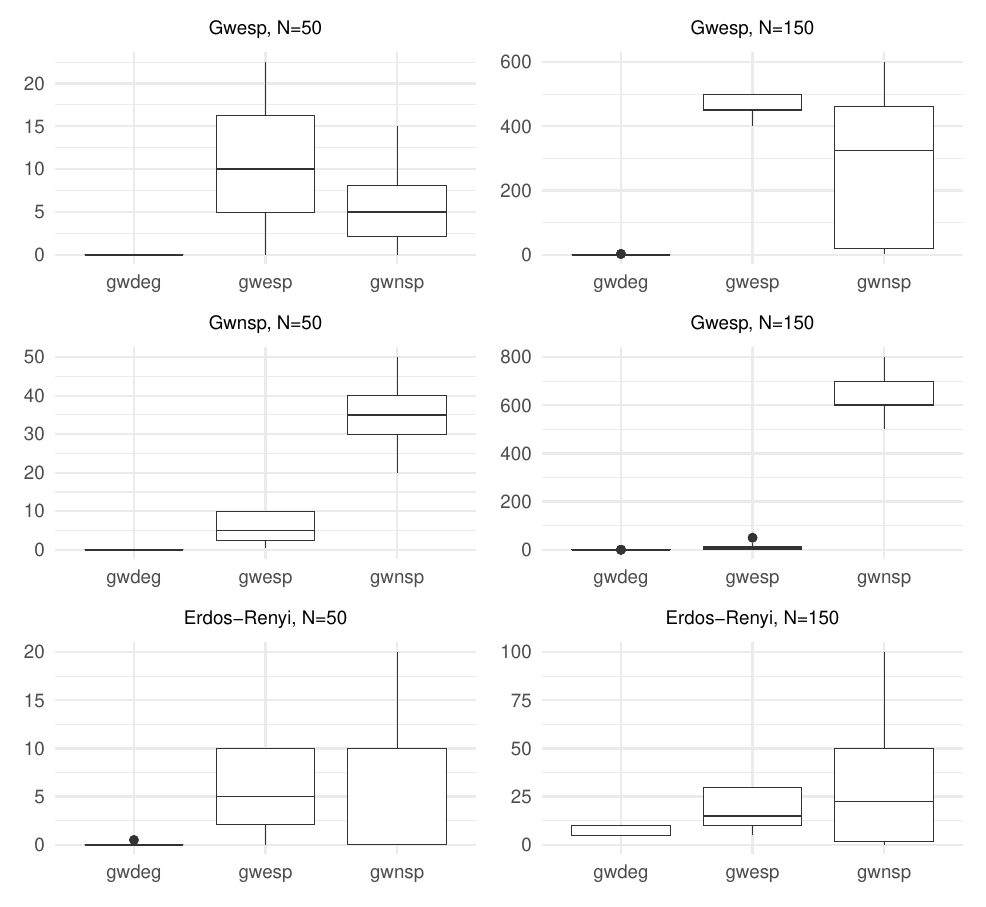}
\caption{Boxplot of the importance scores of each variable in the different simulation setups. The first row corresponds to setup 1, the second to setup 2, the third to setup 3. Left column corresponds to network size $N=50$, right column to network size $N=150$.}
\label{box_simul_plot}
\end{figure}

Figure \ref{simul_plot} summarizes the coefficient values for different penalizations in each different simulation setup: different colors represent the coefficients of different variables. The middle colored line represents the average of each coefficient, over the sample of the 20 networks used in the setup, while the shaded area spans between maximum and minimum. Figure \ref{box_simul_plot} summarizes how importance scores of variables change in different setups. The boxplots highlight how, in setups where a certain variable is used to simulate the data, that same variable tends to have a higher importance score. Notably, in setups where the data is simulated from an edge-independence distribution, no clear distinction in the importance scores of \texttt{gwesp} and \texttt{gwnsp} is detectable and, moreover, an overall lower importance score for all variables is noticeable (especially for larger network size).


\subsection{Endogenous and Exogenous Network Statistics}
The above simulation study concerns endogenous network statistics. If additional exogenous quantities are available, a typical research question when analysing real life social networks is whether the reason behind the tie formation process is driven endogenously or exogenously, i.e.\ depending on the local structure of the ties surrounding the actors or on the actors' intrinsic characteristics. 
Our second simulation study focuses on the capability of LASSO as a variable selection routine to distinguish between endogenous and exogenous variables. As done before, we investigate 3 possible setups and two possible network sizes ($N=50$ and $N=150$). Every network is given a binary nodal attribute, that will influence the formation of edges differently, depending on the simulation setup. In setup 1 the probability of forming ties only depends on the attribute of the connected nodes, and the probability of an edge is set at  $0.05,\ 0.15$ or $0.30$ depending if none, one or both of the nodes have the attribute set to $1$. In setup 2 the probability of forming edges only depends on the \texttt{gwesp} statistic, and the nodal attribute is ignored, while in setup 3 it depends on a combination of both. In each run a sample of 20 networks is drawn.

For the set of statistics to select from, the same endogenous statistics are used as in the previous study, with the addition of the nodal attribute. The nodal statistic, implemented through the \texttt{nodefactor} term in the ERGM package, counts the number of occurrences of a node with a given attribute value in the network edges and is formally defined as
\[
s_{attr}(y) = \sum_i^n\sum_{j<i} y_{i,j} (x_i + x_j),
\]
where $x_{attr}(i)$ is equal to 1 if node $i$ has its binary attribute set to 1 and equal to 0 otherwise.

\begin{figure}
\centering
\includegraphics[width=1.2\textwidth,
height=1.2\textheight,keepaspectratio]{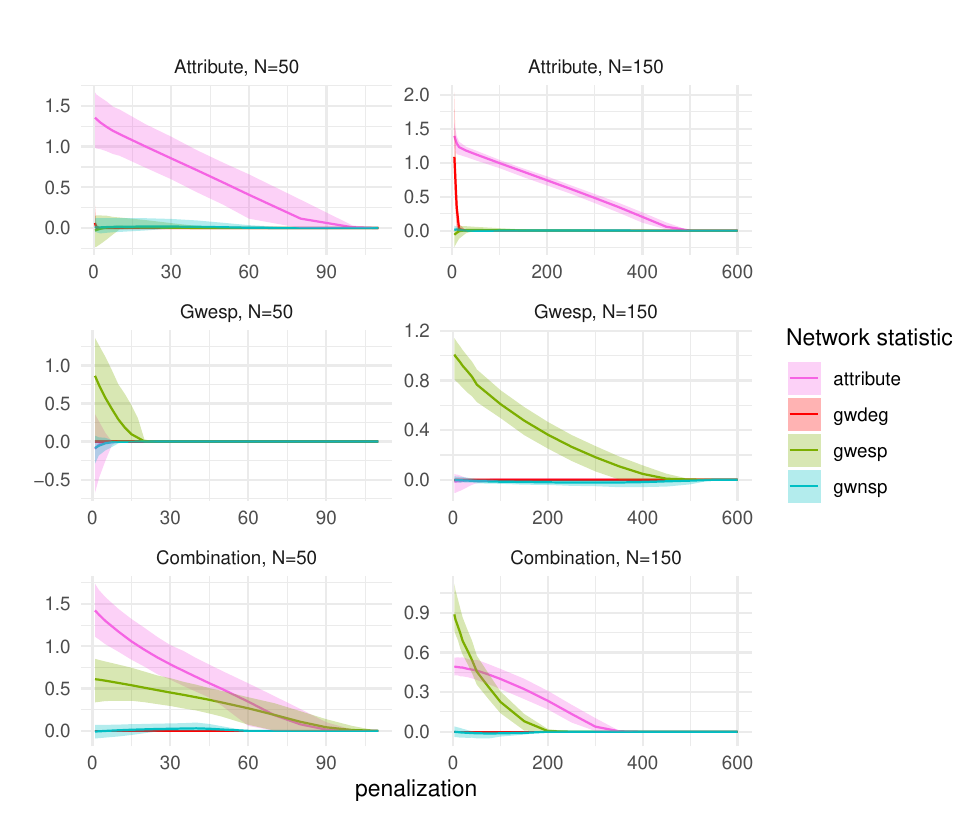}
\caption{Coefficient path in the two simulation setups using a nodal attribute, for decreasing values of the penalization parameter. The first row corresponds to setup 1, the second to setup 2, the third to setup 3. Left column corresponds to network size $N=50$, right column to network size $N=150$}
\label{example3_plots}
\end{figure}

\begin{figure}
\centering
\includegraphics[width=\textwidth,
height=\textheight,keepaspectratio]{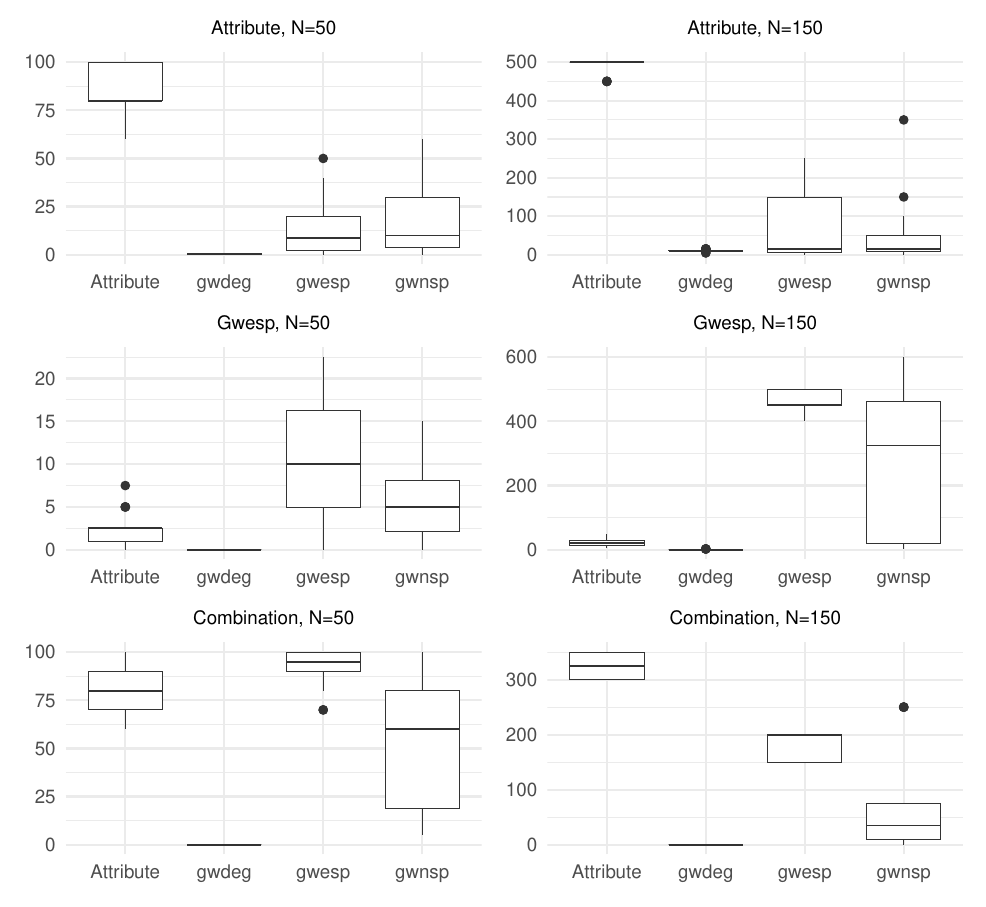}
\caption{Boxplot of the importance scores of each variable in the different simulation setups. The first row corresponds to setup 1, the second to setup 2, the third to setup 3. Left column corresponds to network size $N=50$, right column to network size $N=150$.}
\label{box_attr_plot}
\end{figure}

Figure \ref{example3_plots} and \ref{box_attr_plot} once again summarize the coefficient values for different penalizations in each different simulation setup and how importance scores of variables change in different setups. In setups where the nodal attribute is used to simulate the edges, its importance score tends to be high. On the other hand, in setups where the nodal attribute is bypassed, its importance score plummets. 

\section{Real data applications}

\subsection{London street gang}
\begin{figure}
\centering
\includegraphics[
  height=9.5cm,
  width=\textwidth,keepaspectratio]{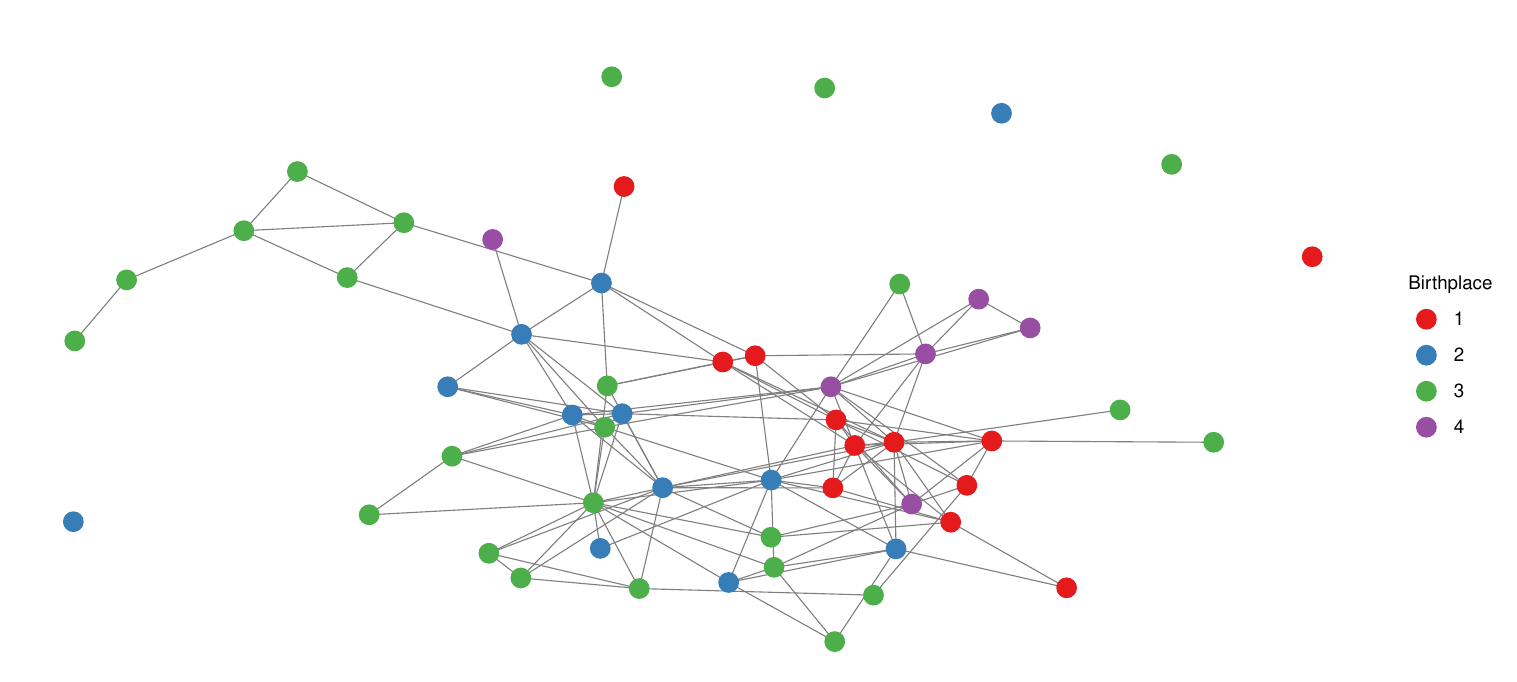}
\caption{London gangs network. Birthplace encoding: 1 = West Africa, 2 = Jamaica, 3 = UK, 4 = Somalia.}
\label{london_gang}
\end{figure}

The first real network we investigate is the \textit{London Street Gang} network, see Figure \ref{london_gang}, first studied by \citet{londongang} and re-used by \citet{taperedergm} in the context of ERGM literature. The 54 nodes in the network represent each a known member of a criminal gang based in London between 2006 and 2009. An undirected tie between two nodes is present if the two gang members have been arrested together while committing a crime. The set of nodal covariates includes age and birthplace of the subject, total number of arrests and convictions, and binary variables representing whether or not the subject has been incarcerated before, is holding a legal residence permit and has been active in the music industry. 

All gang members are ethnically identifiable as black, but their birthplace splits them into 4 different cultural subgroups: West African, Jamaican, UK born and Somali. Part of the goal of the original study by \citet{londongang} was to show that the cultural group had a major influence on how ties are formed between gang members. 

Our goal is to utilize our variable selection procedure to obtain a ranking of importance between the different variables and get an insight into whether this particular network is exogenously or endogenously driven.

Nodal attributes are encoded using the standard terms used in the \textit{ERGM} \texttt{R} package: for categorical attributes one network statistic is added for each level minus one, counting the number of times a node with the set level appears in an edge; for numerical attributes, one single statistic is added, equal to the sum of the attribute value of the connected nodes, overall network edges; the node matching statistics of a given categorical attribute simply count all edges of the network connecting nodes with the same level. More precisely:

\begin{align*}
    s_{numerical}(y,x) &= \sum_{i,j} y_{i,j}(x_i + x_j)\\
    s_{categorical}(y,x = 1) &= \sum_{i,j} y_{i,j}(\mathds{1}(x_i = 1) + \mathds{1}(x_j = 1))\\
    s_{matching}(y,x) &= \sum_{i,j} y_{i,j}(\mathds{1}(x_i = x_j))
\end{align*}


The nodal covariates we include are all the aforementioned individual information on the subjects, while the set of structural statistics, following what was done in the simulation studies, includes the \texttt{gwesp} and \texttt{gwnsp} statistic to measure triadic effects, the \texttt{gwdegree} statistic to measure degree-based effects.



\begin{figure}
\centering
\includegraphics[
  height=9.5cm,
  width=\textwidth,keepaspectratio]{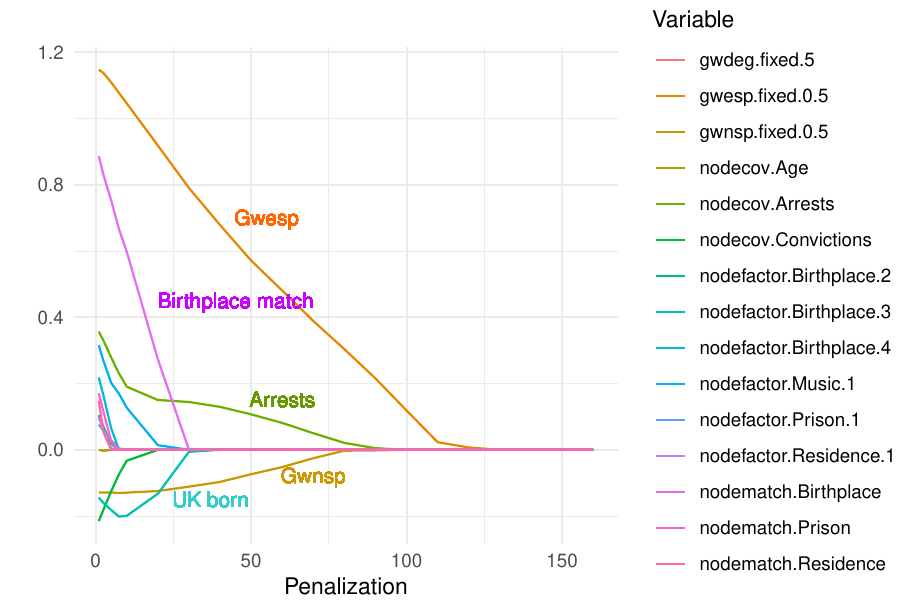}
\caption{Coefficient values given different penalization parameters for the \textit{London Street Gang} network.}
\label{london_gang_plots}
\end{figure}

We show the coefficient paths in Figure \ref{london_gang_plots}.
The results give a valuable insight on tie formation: the network is classifiable as structure-driven, as the edge- and non-edge-wise shared partners statistics are selected earlier than all nodal statistics. The positive and negative signs for the \texttt{gwesp} and \texttt{gwnsp} coefficients, respectively, are a strong indicator of triadic closure, namely the tendency for the edges to form closed triangular structures. 

Of all the nodal covariates, the number of arrests is the first being selected based on the importance score and the only one with an importance value comparable to the structural variables. Other nodal covariates require a lower penalization to be included in the model and thus can be considered of lower importance to the tie formation process. Among the dummy variables encoding the birthplace covariate, it's worthwhile noting the higher importance value for \texttt{Birthplace.3} (UK born). This shows a higher relevance in the tie forming process of being born in the UK, compared to different birthplaces. In addition to the higher relevance, the negative value for the coefficient highlights a negative influence or, in other words, a higher probability for UK-born gang members to occupy marginal roles in the gang. The next variables to get selected are \texttt{nodematch.Birthplace}, indicating a tendency towards associating with gang members born in the same region, \texttt{nodefactor.Music} and \texttt{nodecov.Convictions}. The final model, refit without penalization using the aforementioned variables, achieves an AIC score of 701.50, while adding the next ranked variable (\texttt{nodematch.Prison}) increases the score to 702.35. 

The coefficient estimations for the final model are shown in Table \ref{table_model_london}. 

\bigskip

\begin{table}[h!]
\centering
\begin{tabular}{|p{4cm}||p{2.0cm}|p{2.0cm}|p{2.0cm}|}
 \hline
 Variable& Coefficient & SE & P-value\\
 \hline
 edges   & -3.6440  &  0.3984   &  < 0.0001\\
 gwesp.fixed.0.5 & 1.1575  &  0.1745 & < 0.0001\\
 nodecov.Arrests  &  0.3859  &  0.1191 & 0.0012 \\
 gwnsp.fixed.0.5& -0.1271  &  0.0403   & 0.0016\\
 nodematch.Birthplace & 0.8433  &  0.1352 & < 0.0001 \\
 nodefactor.UK.1 & -0.2019  &  0.0729 & 0.0056 \\
 nodefactor.Music.1  &  0.2408  &  0.1217 & 0.0479 \\
 nodecov.Convictions  &  -0.2255  &  0.1119  &  0.0439 \\
 \hline
\end{tabular}
\caption{Estimated coefficents, standard errors and P-values of the final proposed model for the London Street Gang network.}
\label{table_model_london}
\end{table}

\subsection{Coworker relationship in a corporate law partnership}

The second dataset we apply our method to is the strong coworker network used in \citet{lazega} and later on featured in \citet{SniPatRobHan2006}. The 71 nodes in the network represent lawyers of a Northeastern US corporate law firm, referred to as SG\&R, operative in New England between 1988-1991.
A tie between two members of the firm exists if in the past year, the two lawyers either worked together on a case or an administrative task or exchanged work products. A number of nodal attributes are available, such as gender, age, years spent within the firm, status in the firm (i.e. partner or associate), the office they are based in (Boston, Hartford or Providence), the type of their practice (corporate or litigation) and the law school they went to.

\begin{figure}
\centering
\includegraphics[
  height=\textheight,
  width=\textwidth,keepaspectratio]{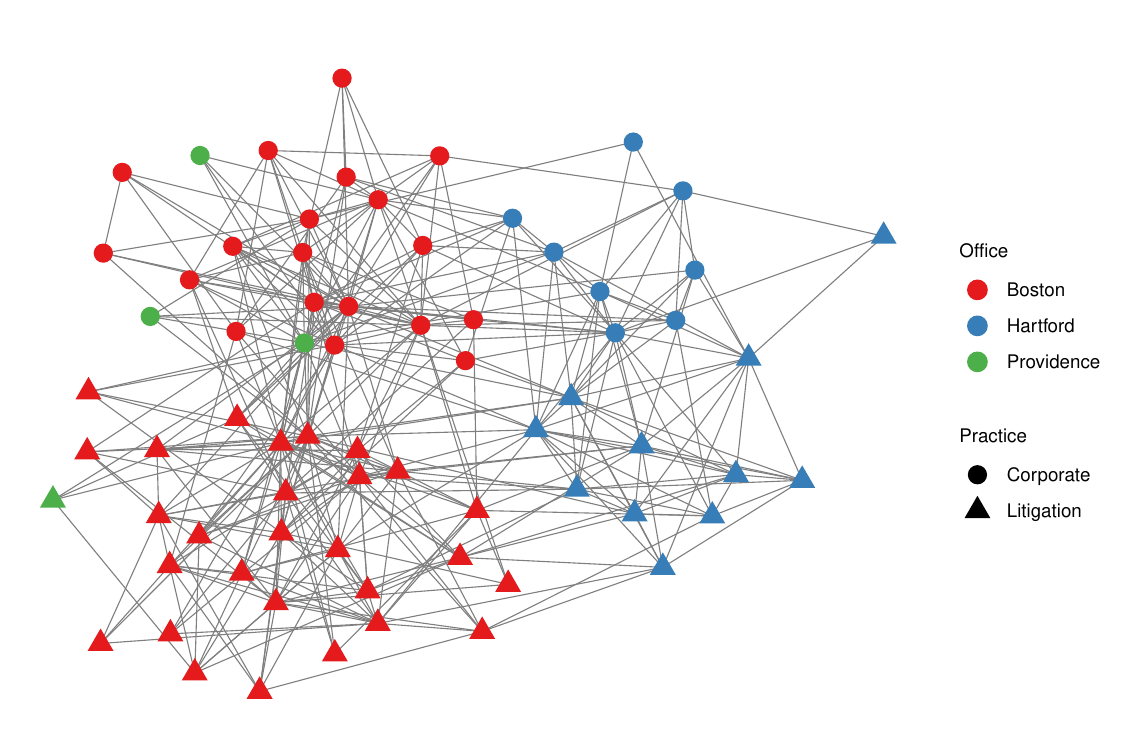}
\caption{Law Firm Coworkers network. Nodal features highlighted: office (colour of the node) and practice (shape of the node).}
\label{example3}
\end{figure}

\begin{figure}
\centering
\includegraphics[
  height=9.5cm,
  width=\textwidth,keepaspectratio]{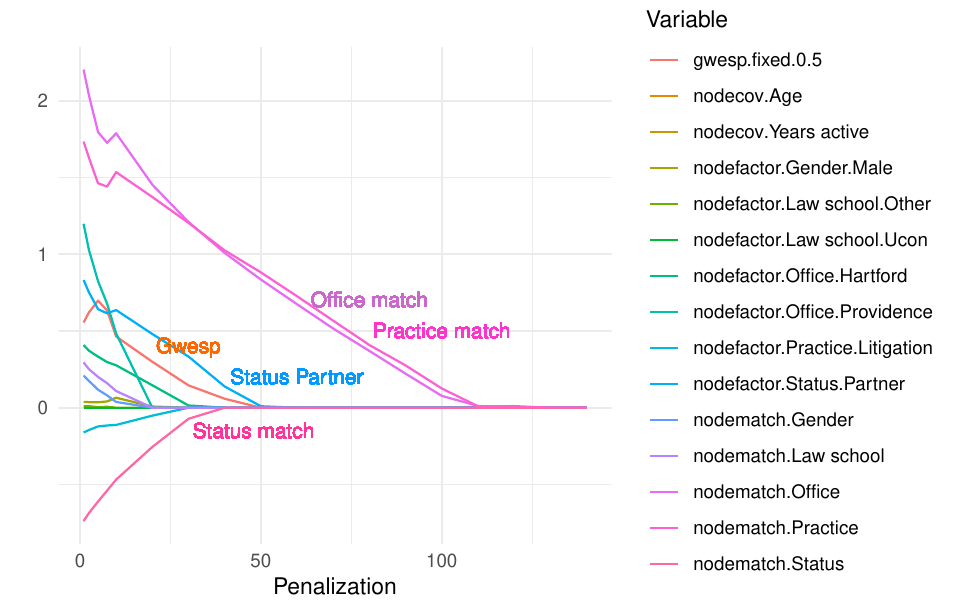}
\caption{Coefficient values given different penalization parameters  for the \textit{Law Firm Coworkers} network.}
\label{lawyers_plots}
\end{figure}

Once again, our goal is to use variable selection to rank the different structural and exogenous variables. The coefficient paths are depicted in Figure \ref{lawyers_plots}.

As opposed to the previous example, this network can be confidently classified as node-driven. In this case, two node attributes, namely the match between offices and the match between practices, are strongly detected as the main drivers for tie formation. Next selected is a group of variables with approximately the same importance score. The structural variable \texttt{gwesp}, as well as the Status of the lawyer and the match between Status. Both the negative value for the \texttt{nodematch.Status} coefficient and the positive value for \texttt{nodefactor.Status.Partner} highlights the tendency for high-profile partners to not collaborate with each other, but at the same time to form a large number of ties with lower-ranked coworkers. The nodal factor office also belongs to this cluster. The rest of the variables in the importance ranking are closely clustered together with very little difference in score, the highest scoring being the match between law schools and the match between gender. We decided to draw the cutting line here: this model achieves an AIC score of 1610.35, while adding the next ranked variable (years active) ends up increasing the AIC score to 1610.63.

The proposed model, making use of the selected variables, is summarized in Table \ref{table_model_lawyers}.

\begin{table}[h!]
\centering
\begin{tabular}{|p{5cm}||p{2.0cm}|p{2.0cm}|p{2.0cm}|}
 \hline
 Variable& Coefficient & SE & P-value\\
 \hline
edges  &  -6.2358 &   0.3038  &   < 0.0001\\
nodematch.Practice  &  1.3942  &  0.1309   &   < 0.0001\\
nodematch.Office   &  1.8223  &  0.1679   &   < 0.0001\\ 
nodefactor.Status.Partner   &  0.5263  &  0.1024   &   < 0.0001\\ 
gwesp.fixed.0.5   &  0.9462  &  0.1567   &   < 0.0001\\
nodematch.Status   & -0.8007  &  0.1580   &   < 0.0001\\ 
nodefactor.Office.Hartford  &  0.4187  &  0.0922   &   < 0.0001\\
nodefactor.Office.Providence &  0.9854  &  0.1912   &   < 0.0001\\ 
nodematch.Law school  &  0.3203  &  0.1326   &   0.0157  \\
nodematch.Gender  &  0.2368  &  0.1138  &  0.0374\\
 \hline
\end{tabular}
\caption{Estimated coefficents, standard errors and P-values of the final proposed model for the Law Firm Coworkers network.}
\label{table_model_lawyers}
\end{table}

\section{Discussion}

The usage of a penalized likelihood provides for an effective variable selection procedure in the framework of Exponential Random Graph models. Refitting the penalized model for decreasing values of the penalization parameter gives an order of importance for the initial set of variables, where one is less important than another if a smaller penalization is needed to obtain a non-zero coefficient. In other words, the value of the smallest penalization parameter for which a given variable has a non-zero coefficient can be used as an importance score for the variable. As noted before, the coefficients obtained are biased and for this reason, we do not use them for the final model but, instead, refit an unpenalized model using the selected variables.

The importance score can be derived for both structural statistics and nodal attributes making it possible to deduce if the main driver for tie formation in the network is endogenous, in case the structural statistics rank high, or exogenous, if nodal covariates rank higher.

Model selection is achieved through setting a cutoff value for the importance score, and choosing the variables with higher scores than the cutoff. No indication is given on the choice of the cutoff, which can be
tailored to the specific needs of the case, or be done following other model quality metrics, such as AIC or variable p-values.

In the present paper we only deal with static, undirected networks, but as ERGMs or ERG-derived models are widely used to analyse both directed networks and networks that evolve through time. A possible direction for additional development and future work will be to extend the usability of the LASSO based variable selector to both directed and temporal ERGMs.

\printbibliography

\end{document}